# Multiwavelength Observations for a Double-decker Filament Channel in AR 13102

Yin Zhang[1,2], Baolin Tan[1,2,3], Quan Wang[1,2,3], Jing Huang[1,2,3], Zhe Xu[4], Kanfan Ji[4], Xiao Yang[1,2], Jie Chen[1,2], Xianyong Bai[1,2,3], Zhenyong Hou[5], and Yuanyong Deng[1,2,3]
[1] CAS Key Laboratory of Solar Activity, National Astronomical Observatories, Chinese Academy of Sciences, Beijing 100012, People's Republic of China
[2] Key Laboratory of Solar Activity and Space Weather, National Space Science Center, Chinese Academy of Sciences, Beijing 100190, People's Republic of China
[3] School of Astronomy and Space Science, University of Chinese Academy of Sciences, Beijing 100049, People's Republic of China
[4] Yunnan Astronomical Observatories, Chinese Academy of Sciences, Kunming 650216, People's Republic of China
[5] School of Earth and Space Sciences, Peking University, Beijing 100871, People's Republic of China
Received 2023 September 6; revised 2024 June 3; accepted 2024 June 21; published 2024 September 11

## Abstract

We present the observational evidence of the existence of a double-decker filament channel (FC) by using observations in extreme ultraviolet and Hα wavelengths. For both FCs, the east foot-point roots in the active region (AR), while the west one roots in the remote quiet region. The bottom FC (FC1) appears as intermittent filaments. Within the AR, the FC1 appears as an S-shaped filament (F1), which consisted of two J-shaped filaments (F1S/F1N for the south/north one). For the upper one (FC2), only the east part is filled with dark plasma and visible as a small filament (F2). Its east foot-point roots around the junction of F1S and F1N. Initially, due to the recurrent reconnections, F1N and F1S link to each other and form a new filament (F3) thread by thread. Meanwhile, the heated plasma, which appears as brightening features, flows from the east foot-point of F2 to the west, and becomes invisible about $1.1 \times 10^5$ km away. The failed eruption of F1S is triggered by the reconnection, which appears as the brightening threads changing their configuration from crossed to quasiparallel in between the F1S and F3, and is confined by the upper magnetic field. Associated with the eruption, the distant invisible plasma becomes visible as a brightening feature. It continuously flows to the remote foot-point, and becomes invisible before reaching it. The brightening plasma flow outlines the skeleton of FC2 gradually. The observations show the existence of a double-decker FC, as a magnetic structure, before they appear as a brightening/dark feature when fully filled with hot/cool plasma.

*Unified Astronomy Thesaurus concepts:* Solar activity (1475)

*Materials only available in the* online version of record: animation

## 1. Introduction

A solar filament, which is defined as the relatively cool and dense plasma suspended in the solar chromosphere, has been observed in the Hα line for several decades. It is also called a prominence when it appears on the solar limb. Recent observations with extreme ultraviolet (EUV) wavelengths reveal that the filament is more extended with wavelengths below the hydrogen Lyman-continuum edge (912 Å) than in the Hα wavelength (Heinzel et al. 2001; Schmieder et al. 2003, 2004; Schwartz et al. 2012).

The filament usually locates along magnetic polarity inversion lines (PILs; Babcock & Babcock 1955), which are defined as filament channels (FCs). Martin (1990, 1998) summarized the conditions for the formation of a filament and proposed that the PIL is only a necessary, but not a sufficient, condition for filament formation. From an observational point of view, FCs can be identified by relatively long-lived, narrow lanes along the PILs at photospheric heights. In the chromosphere, FCs are regions characterized by the approximately parallel alignment of fibrils along the PIL. An FC may also extend to greater heights; in the low corona, it appears as a diffuse dark corridor (Schmieder et al. 2004). The large FC can build up gradually via reconnection events, which occurs between the filaments or between the filament and other magnetic structures (Li et al. 2022a). The FC only appears when fully filled with dark or brightening plasma, i.e., a filament formed along it (Yang et al. 2021; Zhang et al. 2021). Observational studies and numerical simulations reveal that a large filament can be formed by the interaction between two nearby filaments, located in the same FC with the same chirality or different FCs with the different chirality (Schmieder et al. 2004; van Ballegooijen 2004; Aulanier et al. 2006; Su et al. 2007; Chandra et al. 2011; Jiang et al. 2014; Joshi et al. 2014, 2016). These reconnection processes are reproduced with more detail by Linton et al. (2001) in a simulation work and Gekelman et al. (2012) in a laboratory experiment. More details about filaments and FCs can be found in Vial & Engvold (2015).

Thanks to the development of observing techniques for attaining high-spatial-resolution Hα observations, such as the Swedish Solar Telescope (Scharmer et al. 2003), the Goode Solar Telescope (known as the New Solar Telescope before 2017 July; Cao et al. 2010a), and the New Vacuum Solar Telescope (NVST. Liu et al. 2014), the filaments are well observed in more details. These include the the properties of the threads (Lin et al. 2005, 2009, 2012), the counterstreaming motions along the filament (Wang et al. 2018; Yang et al. 2019), possible material replenishment of the filament (Zou et al. 2016, 2017, 2019), and the upflows around the filament foot-point (Cao et al. 2010b). More progress can be found on the websites of these telescopes. Meanwhile, a variety of evolutionary phenomena of filament threads and the ubiquitous small-scale reconnection between filament threads and other magnetic features are also recorded. By a careful case study, various phenomena, which, expected by numerical simulation or phenomenological models of reconnection, are observed, such as current sheets, bidirectional flows, and reconfiguration of the threads involved in the reconnection, an untwisting motion that represents the release of magnetic energy, bright cusp-shaped structures, and underlying magnetic flux cancellation or emergence (Yan et al. 2014; Xue et al. 2016; Huang et al. 2018; Xue et al. 2020; Zheng et al. 2020; Guo et al. 2021; Fang et al. 2022; Li et al. 2022b; Yang et al. 2023). By investigating a partial eruption of a filament, Bi et al. (2015) found that it is composed of two twisted flux ropes, which wind around each other. Chen et al. (2020) present the detailed evolution of the formation of the longer filament by reconnection between aligned short chromospheric fibrils. Li et al. (2022a) report the FC formation by a series of small-scale reconnection events and appearance as the filament by filament materials transfer to the longer and more twisted magnetic field lines. By studying the interactions between filaments and network fields, Song et al. (2023) proposed a new picture of filament material drainage. Sun et al. (2023) present the observational evidence that the FC is filled by direct material injection due to the small-scale magnetic reconnections. The observations also provide the direct observational evidences for the twist of the filament and its associated magnetic flux rope (Yang et al. 2014; Yan et al. 2020), and give more details about the filament interaction (Yang et al. 2017, 2023).

The double-decker filament, which appears as two filaments lying along the same PIL and is hosted by complex magnetic structure, is first identified by Liu et al. (2012). They proposed that the double-decker filament may be hosted by double flux rope structure or a single flux rope situated above an arcade, which contains the lower filament branch. The first scenario is supported by several observational studies (Cheng et al. 2014; Zhu et al. 2015; Dhakal et al. 2018; Hou et al. 2018). Its dynamic evolution is reconstructed by Kliem et al. (2014). The second scenario is supported by Liu et al. (2018) and Awasthi et al. (2019). Based on the flux rope insertion method, Liu et al. (2018) find that the high-lying structure can be formed by the reconnection between two sheared arcades, and the lowing filament is supported by sheared arcades. Meanwhile, other observations suggest that the double-decker filament can be formed by the splitting of an original single filament (Tian et al. 2018; Monga et al. 2021; Pan et al. 2021; Zhang et al. 2022). Joshi et al. (2020) present a complex eruption, which is begun from the configuration of a stack of three flux ropes above the PIL. These studies reveal that the magnetic topology and plasma behavior above the PIL within the active region (AR) is more complex.

In the present study, we address these issues in analyzing the complex activities of a double-decker FC. Both FCs, which host the filaments, are large-scale magnetic structures, with the east foot-point rooting in the core region of AR 13102, and the west one rooting in the remote quiet region. The bottom FC (FC1) is visible as a long intermittent filament. The part located around the east foot-point and within the core region of AR 13102 appears as an S-shaped filament. The S-shaped filament is formed by two small J-shaped filaments. For the upper one (FC2), only the part located within the core region of AR 13102 is visible as a small filament, with the foot-point located around the junction of the J-shaped filaments. The activities, such as recurrent reconnection and the failed filament eruption, mainly occur within the core region. Meanwhile, the associated observational feature, which appears as hot plasma flow, highlights the large-scale magnetic structure of FC2. This observational evidence provide some unique insight into the nature of solar FC. The paper is organized as follows: The data reduction are described in Section 2. The observational results are described in Section 3. The conclusions are given in Section 4.

## 2. Data Source and Reduction

The Solar Dynamics Observatory (SDO) is a comprehensive solar dedicated satellite. The Atmospheric Imaging Assembly (AIA; Lemen et al. 2012) on board the SDO provides full-disk solar images in ultraviolet (UV) and EUV wavelengths with the pixel size and temporal resolution of $0.''6$ and 12 s. Two UV bandpasses focus on the photosphere, with 1700 Å for temperature minimum (5000 K) and 1600 Å for the upper photosphere and transition region (0.1 MK). Another six EUV bandpasses of the AIA are 171, 193, 211, 335, 94, and 131 Å, with the temperature range from 0.6 to 10.0 MK. Their targets are the transition region, corona, and flaring region. The Solar Upper Transition Region Imager (SUTRI) on board the Space Advanced Technology demonstration satellite (SATech-01) provides another EUV wavelength with 465 Å (Bai et al. 2023). It adds a point in the temperature sequence between the AIA UV and EUV bandpasses, which focus on the middle corona and corresponds to about 0.5 MK (Tian 2017). The time cadence and pixel size of SUTRI is about 30 s and $1.''22$. In the present study, the multiwavelength observations combined by AIA and SUTRI record the temporal evolution of the large-scale plasma flow and the confined filament eruption. In order to show the whole profile of FC2 whose different sections were filled with brightening plasma flow at different moments, a composite image is used in the present work. In this composite image, the value of each pixel is the maximum value during a special time period. We label them as a maximum value image in the following sections.

The failed filament eruption is observed by the NVST during the time period from 01:52 to 03:17 UT. The NVST observes the Sun around two wavelengths of TiO and Hα, with a 986 mm clear aperture. In the present study, the Hα images recorded at line center (6562.8 Å) and two off-bands (±0.4 Å) are used. These images are reconstructed to Level 1+ data by the speckle masking method (Weigelt 1977; Lohmann et al. 1983; Xiang et al. 2016). The Level 1+ NVST data can be accessed at the Fuxian Lake Solar Observatory website, with the dark current subtracted and flat-field corrected. The spatial pixel size of these data is about $0.''165$, and the time cadence is about 43 s. The different off-band and line-center images are coalignment by the method proposed by Ji et al. (2019) and Cai et al. (2022). In the present paper, these images are shown as a pseudocolor image, with red for Hα red wing, green for line center, and blue for blue wing. The NVST and AIA data are coaligned by using the cross-correlation technique with the same feature of the bright flare ribbons. Since the field of view of NVST is about $172'' \times 172''$, the full-disk Hα images provided by Huairou Solar Observing Station (HSOS) are also used to show the large-scale morphology of the filament in Hα.

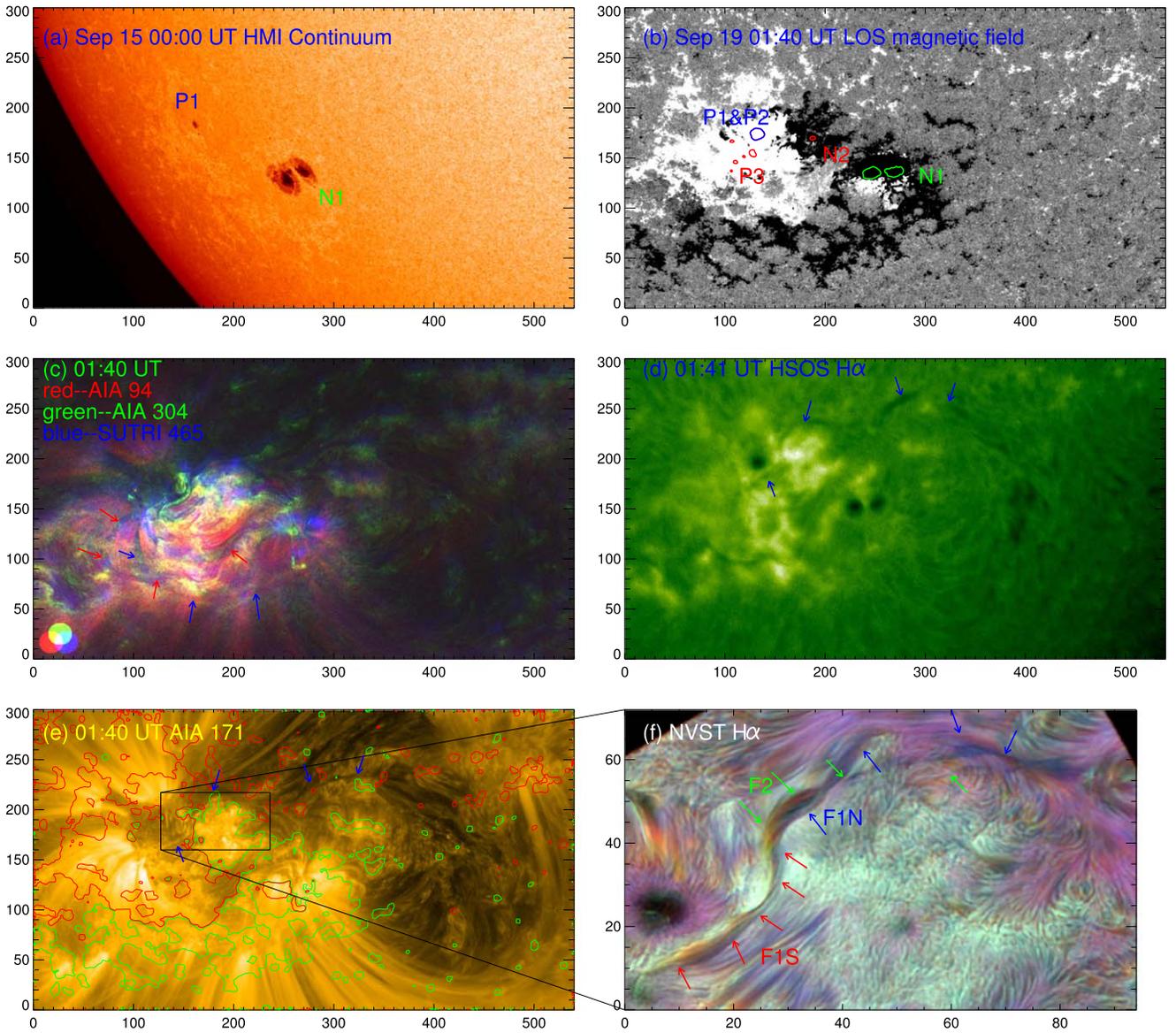

**Figure 1.** Large-scale morphology prior to the eruption. (a) HMI continuum image on September 15 around 00:00. It shows the morphology of AR 13102 in its early evolution phase. (b) HMI line-of-sight magnetic field, with black for negative and white for positive polarities. Contours outline the sunspot's umbra area with intensity smaller than 0.7 times the average intensity in quiet region in continuum image. (c) Composite image with red for 94 Å, green for 304 Å, and blue for 465 Å. Red and blue arrows outline the profile of the large-scale connections. (d) Hα image recorded by HSOS. Blue arrows outline the profile of FC1; (e) SDO/AIA 171 Å image, with green (red) contours representing negative (positive) magnetic polarities (±50 G). (f) High-spatial-resolution Hα image obtained from NVST.

## 3. Observational Results

### 3.1. Overview and the Existence of Bottom FC1

According to the temporal evolution, the AR 13102 experiences at least two emerging processes until the activities we investigated in the present study. On 2022 September 13, when it first appears on the east solar disk, it is a bipolar sunspot formed by very disperse following positive polarities (P1) and strong leading negative spots (N1). Its morphology at the early stage is shown in Figure 1(a) with a Helioseismic and Magnetic Imager (HMI) continuum image recorded around September 15 00:00 UT. On September 15 around 20:00 UT, new opposite polarities emerge in the disperse positive polarity area. Associated with the emergence, the morphology of AR 13102 changes a lot. Its morphology on September 19 is shown in Figures 1(b)–(e). Figure 1(b) is the line-of-sight (LOS) magnetic field. The contours of different colors outline the sunspot umbra in continuum image where the intensity is <0.7 times the average intensity in the quiet region. Comparing Figures 1(a) with (b), the newly emerged positive polarities gather in two places. One is the preexisting positive sunspot. The newly emerged positive polarities merge with the preexisting one. They rotate around each other to form the main positive polarity as circled by a blue contour and labeled as P1 and P2. The others move to the southeast fast, and remain stable on the south of the previous positive sunspot (indicated by a red contour and labeled as P3). On the contrary, the negative polarities cancel with the preexisting positive polarities and form a disperse large area. Their center sunspot umbra is indicated by a red contour and labeled as N2. Several small positive sunspots also can be identified on the east of P1 and P2, and P3. Since they are not involved in the activities we studied in the present work, we ignore them in the present study. Figure 1(c) is a composite image with red for 94 Å, green for 304 Å, and blue for 465 Å. In this image, large-scale connections between the far away opposite polarities can be identified easily in the multitemperature structure as indicated by red and blue arrows. Figure 1(d) is Hα image recorded by HSOS. Comparing with the LOS magnetic field (Figure 1(b)) and Hα image, the long FC1 is guided by blue arrows. The FC1 appears as intermittent filament structures and extends along the large-scale PIL from within the newly emerging area of AR 13102 to the weak magnetic field area on the west to AR 13102. Figure 1(e) is AIA 171 Å image, with red/green circled as the areas with magnetic field strength greater than ±50 G. The blue arrows indicate the continuous swaths of dark features, which is suggested as the counterpart of the FC1 in Hα.

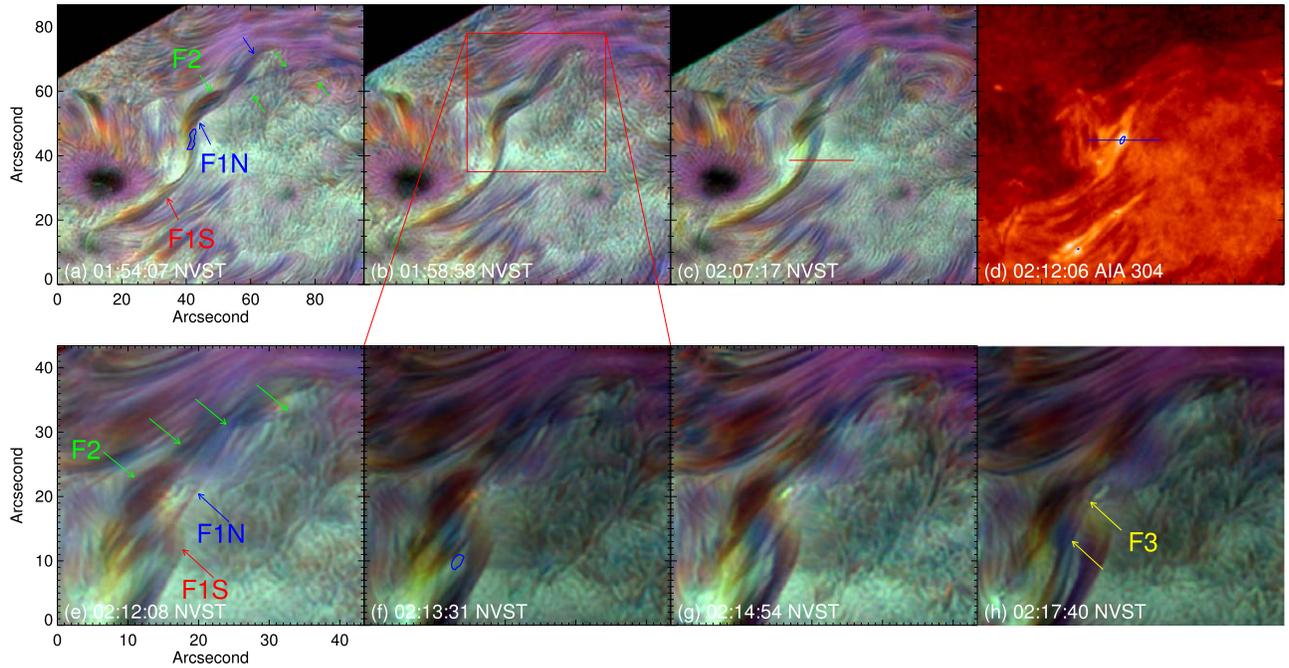

**Figure 2.** Dynamic evolution of the preeruption reconnection. Images in panels (a), (b), (c), (e), (f), (g), and (h) are Hα images, which are recorded by NVST. Panel (d) is an AIA 304 Å image. Images in the first row show the whole morphology of F1S and F1N. Images in the second row are an enlarged image, outlined by red box in panel (b). The filaments of F1N, F1S, F2, and F3 are indicated by blue, red, green, and yellow arrows. Blue contours outline the EUV brightening recorded in 304 Å. Red and blue lines in panels (c) and (d) show the position of the slit, shown in Figure 3.

As outlined by black box in Figure 1(e), the small filament part, located along the PIL between N2, and P1 and P2, is shown in Figure 1(f) with enlarged Hα image recorded by NVST. The filament fine structures, which made up by many thin threads extending along the filament spine, can be identified clearly in NVST observations. Due to the projection effect, the double-decker filament is partially overlapped. As shown in Figure 1(f), the S-shaped bottom filament is indicated by red/blue arrows for south (F1S)/north (F1N) part. The morphological characteristics of the F1S are that several quasiparallel threads originate from its south leg, draw a small arc, and then squeeze together around its north leg. The threads in F1N draw an inverse arc, and locate along the PIL head to tail with the F1S to form the S-shaped structure. The S-shaped filament is part of the FC1 as we outlined in Figure 1(d). The profile of the twisted upper filament (F2) is outlined by blue arrows. It is overlapped with F1N and can only be identified during the process of the plasma flow as shown in Figure 1(f). In this image, F2 appears as a dark leg overlapped with F1N and a dispersion body of the plasma flow that formed by several threads. These dark features outline the profile of a transmeridional FC, which is labeled as FC2.

### 3.2. Appearance of Upper FC2

#### 3.2.1. Recurrent Reconnections

As shown in Figure 1(f), three nearby foot-points, such as the east foot-point of F2, the north foot-point of F1S, and the south foot-point of F1N, are rooted in the adjacent area. According to multiwavelength observations from AIA, the recurrent small-scale reconnections, which appear as intermittent EUV brightening, start around 01:42 UT. The NVST observation starts from 01:52 UT. It shows that the EUV brightening is cospatial and cotemporal with the interaction between threads of F1S and F1N, and the following reconfiguration of these threads.

The NVST Hα images in Figures 2(a)–(c) show the morphology evolution of F1S and F1N associated with reconnections at three selected times. A blue contour outlines the EUV brightening in 304 Å image where the intensity is >0.6 times the maximum intensity in AR. In Figure 2(a), F2, F1S, and F1N are indicated by green, red, and blue arrows, respectively. The EUV brightening locates in between the south foot-point of F1N and the north foot-point of F1S. As shown in Figure 2(a), the F1S is composed by several highly sheared Hα threads. Associated with the reconnection, the F1S evolved into a thin and compact small filament as shown in Figure 2(b) and then into loosely aligned threads as shown in Figure 2(c). Figures 2(d)–(h) show the detail of the reconfiguration of F1S and F1N thread by thread during the reconnection process, which starts at 02:11 UT and ends around 02:17 UT. The associated EUV brightening, which can be identified in all AIA observational wavelengths, is initial around 02:11 UT as shown in Figure 2(d) with AIA 304 Å image. The enlarged image in the second row is NVST Hα images for the core region as indicated by red box in Figure 2(b). The brightening in 304 Å image as outlined by a blue contour is overlaid on Figure 2(f). It shows that the EUV brightening is located around the intersectional place between the threads, which belong to F1S and F1N. As shown in Figures 2(e)–(g), the threads in F1N adjacent to the intersectional point connect smoothly to the threads in F1S. During this period, the threads of F1S become loose in the reconnection region and then reconnect with F1N in turn. Subsequently, part of F1N and F1S merges and becomes a little bit longer filament (F3), as indicated by yellow arrows in Figure 2(h). During the time period from 01:42 to 02:17 UT, the EUV brightening locates around the interactions area between F1N and F1S, which appears as the signature of reconnection and results in the reconfiguration of the threads.

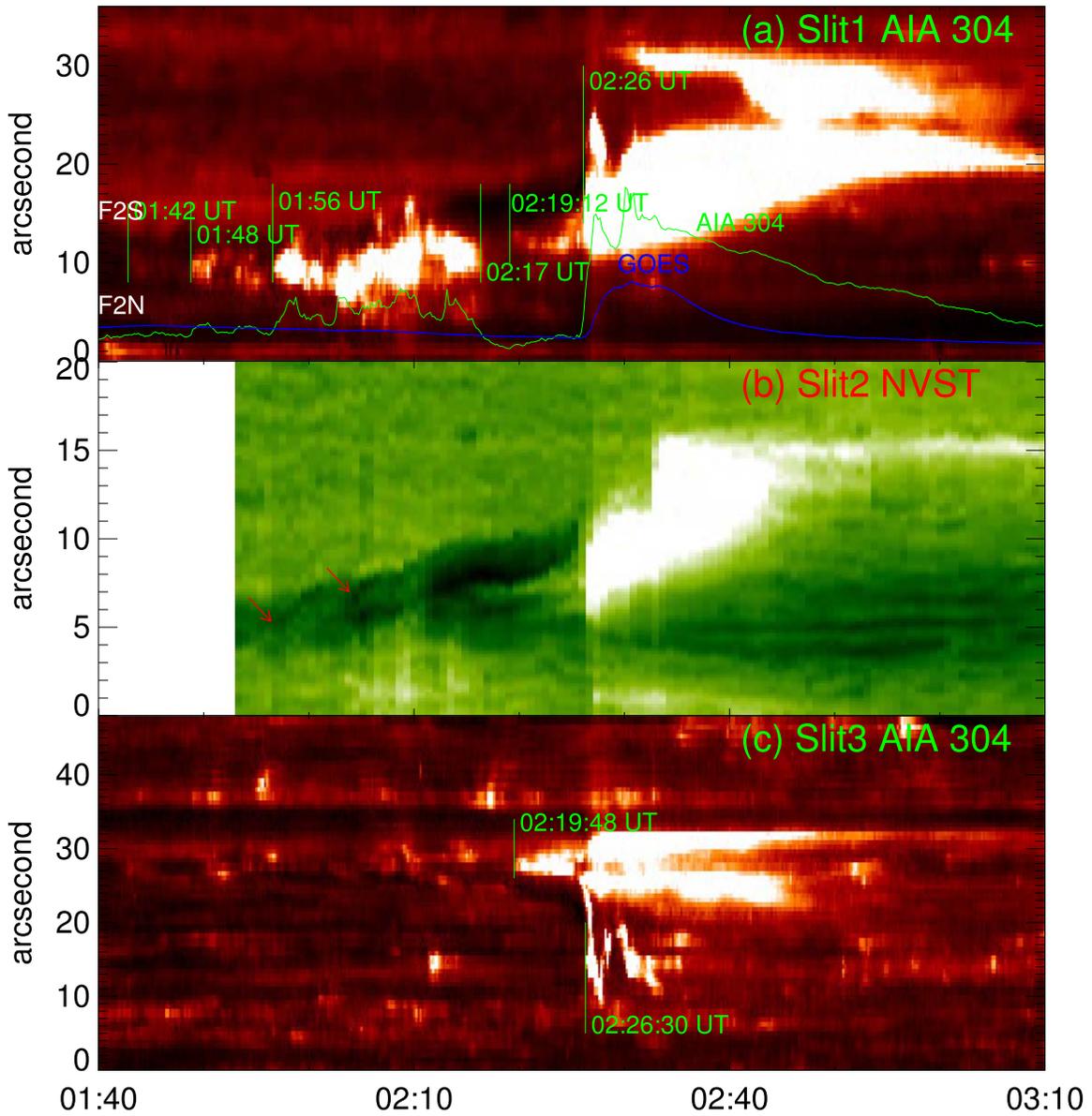

**Figure 3.** Time-slice plot acquired along the slit as shown in Figures 2(c), (d), and 4(d).

Here, we show a time–distance diagram of 304 Å image in Figure 3(a), which shows the evolution of EUV brightening smoothly and obviously. The green curve is the normalized integral intensity along the slit. The blue curve shows the normalized value of GOES 1–8 Å flux. Each small-scale reconnection corresponds to a small bump in the light curve of all AIA channel (green curve of 304 Å as an example). The horizontal blue line (Slit1) in Figure 2(d) crosses the reconnection region. Along this line, the reconnection feature, which appeared as the EUV brightening, is almost the same in all EUV wavelengths. The horizontal red line in Figure 2(c) shows the position of Slit2. It crosses the north part of F1S, but misses the reconnection area. The dynamic evolution of the threads seen through Slit2 is shown in Figure 3(b). In both diagrams, the F1N and F1S appear as the dark features. The brightening as a signal of the reconnection appears intermittently with the staircase-like intensity increase. As shown in Figure 3(a), the first EUV/UV brightening is recorded around 01:42 UT with a very small increments. Around 01:48 UT, with a new reconnection occurring, the intensity of the EUV brightening increases. This process lasts for about 8 minutes and ends around 01:56 UT. The intense EUV brightening starts again at almost the same time. It lasts about 20 minutes and ends around 02:17 UT. Associated with the EUV brightening, the untwist of the thread is identified as the movement as shown in Figure 3(b). The green arrows indicate two threads. The first one is the only recognizable thread that moves from east to west (seen from the solar disk). The second one appears when the first one disappeared at the west edge. Another thick thread appears at the same time with the second one. It also moves from east to west. The tracks of the threads seen through the slit are a manifestation of the emergence of a helical structure.

The mergence of two filaments, located along the same FC, are reported by several previous studies. It usually associates with the reconnection between the magnetic structure, which hosted the filaments and resulted in a flare (Schmieder et al. 2004; Kumar et al. 2010). van Ballegooijen (2004) presented an observation of the mergence of two individual filaments, hosted by the same magnetic flux rope, but without any brightening in observational wavelengths. Chandra et al. (2011) studied the same case as Kumar et al. (2010). According to the observational evidence that the flare-associated brightenings, observed in the available wavelengths, are not cospatial with the the merging point of the filaments, Chandra et al. (2011) suggested that there was no causal relationship between the mergence of the filaments and the subsequent flare. The mergence of filaments, located in different FC, is presented by Su et al. (2007). Several recent observations further showed that, during the rising phase, the filament can partially merge with other filament staying in its way (Bi et al. 2012; Liu et al. 2012). Jiang et al. (2014) presented a mergence that occurs between an erupted filament and an FC standing in its way, in which another filament was embedded. Under the convective-zone conditions, Linton et al. (2001) simulated four fundamental types of the interactions between two highly twisted and identical flux ropes. Linton (2006) further presented the reconnection of flux tubes with low twist and nonequal axial fluxes. Considering the filaments as a structure, which is supported by sheared arcade or magnetic flux rope in the solar atmosphere, Török et al. (2011) performed the 3D MHD numerical simulation under the corona conditions with two filaments changing their connectivity to form two newly linked filaments. It can explain the filament interaction observed by Kumar et al. (2010) and Chandra et al. (2011) successfully. Here, in our event, the mergence occurs between the filament threads within the same FC. The EUV and Hα brightening, which can be regarded as the signal of reconnection, are observed. Meanwhile, the reconnection has no direct relationship with the following filament eruption and flare, which we will discuss in more detail in the following section.

### 3.2.2. Failed Filament Eruption

The eruption of the remained F1S is triggered by the reconnection between threads in between F1S and F3 and shown in Figure 4. In order to highlight the key morphology at different times, the images at different wavelengths are used. The reconnection is initial as the threads brightening, which first appears around 02:19 UT. Figure 4(a) shows the morphology of the EUV brightening in a 94 Å image. This brightening can be identified in all AIA wavelengths, but more clear in 94 Å. As shown in the 94 Å image, the crossed threads in between remain F1S and F3 become brightening simultaneously, with the morphology outlined by red and yellow lines in a subimage at the bottom right of Figure 4(a). The brightening feature cannot be identified in the Hα image as shown in Figures 4(b) and (c). Meanwhile, in the high-spatial-resolution image, the departure of F1S (red arrows) and stability of F3 (yellow arrows) is recorded. The reconnection occurs around the cross of the threads, and the reconfiguration can be seen clearly in 94 Å image as shown in Figures 4(d). Red and yellow arrows indicated the parallel brightening threads. Associated with the reconnection, the F1S moves outward and disappears in the Hα image from about 02:25:59 UT (Figures 4(e)). In AIA EUV images, the upward-moving dark filament becomes a brightening around 02:26:40 UT and reaches its maximum height around 02:27 (Figures 4(g)). Then, it shrinks as associated with the disappearance of the top part; see Figures 4(h)–(i). As shown in Figure 4(i), the top part of the erupted filament is too weak to be identified. The whole erupted filament structure reappears around 02:30 UT as shown in Figure 4(j). Meanwhile, the whole structure of the loop structures located above the erupted structure appears around that time as indicated by red arrows. Tracking back the leg of the loop structure, it can be identified as early as around 02:26 UT, which is consistent with the reconfiguration of the brightening threads. As shown in the following images of Figures 4(k)–(l), the loop structures remain stable and always located above the erupted F1S. Meanwhile, their foot-points are rooted along the brightening ribbon as we had seen in the Hα image. This observational evidence show that the erupted filament is confined by the upper loops, and the two ribbons are the foot-print of these loops.

Figure 3(c) shows the time–distance diagrams at the top of the filament with the slits as indicated by the white line in Figure 4(d). As shown in Figure 3(c), the sudden eruption of the filament begins with the EUV brightening under the filament and the associated slow rising of the filament. This EUV brightening first appears under the filament around 02:19 UT. It also can be identified in Figure 3(a) at the same time, which is consistent with the description above. Then, the F1S suddenly erupts around 02:25:42 UT and changes into a brightening feature around 02:26:30 UT.

By the nonlinear force free magnetic field (NLFFF; Wiegelmann 2004) and a potential field extrapolation, we reconstruct the coronal magnetic field based on the HMI vector magnetic field. The magnetic field is calculated in a Cartesian box of $140 \times 115 \times 100$ arcsec$^3$. The parameter-decay index, which is a quantitative description of how fast the closed potential magnetic field decreased with height, is calculated and shown in Figure 5. Following Sun et al. (2015), the flaring polarity inversion line (FPIL) mask, where most free energy is stored, and which is used to demarcate the AR core field, is designed as follows. We isolate the PIL from the LOS magnetic field map, and dilate them with a circular kernel (radius $r = 2.2$ Mm). It is shown in Figure 5(a) as the blue contour. An AIA 1600 Å image, which was recorded around the flare peak, is used to obtain the flaring region. The flaring region is defined as the pixel with the intensity above 700 digital numbers per second, and shown as a red contour. As shown in Figure 5(a), the flare ribbons of this event are not parallel with the PIL. It is consistent with the previous observational evidence, as we show in Section 3.2, that the brightening ribbons are the foot-points of coronal loops, located above the failed erupted filament. So, here, we use the dilated PIL within the flaring region as our FPIL mask (yellow contour). For the present event, the initial active height of the filament is hard to determine. A snapshot of the NLFFF extrapolated magnetic field in X–Z plane along the green line in Figure 5(a) is shown in Figure 5(b). Here, the x-axis is along the east–west direction, the y-axis is along the north–south direction, and the Z is the height from the solar surface, in units of megameters. A twisted structure is easy to identify. We suggest that it is the counterpart of the filament in the coronal magnetic field. The height of its top edge, as indicated by a green line, is around 10 Mm. The variation of the decay index along the height is shown in Figure 5(b). In the scatter plot, $n$ is about 0.47 around 10 Mm, which is much smaller than the common threshold value of 1.5.

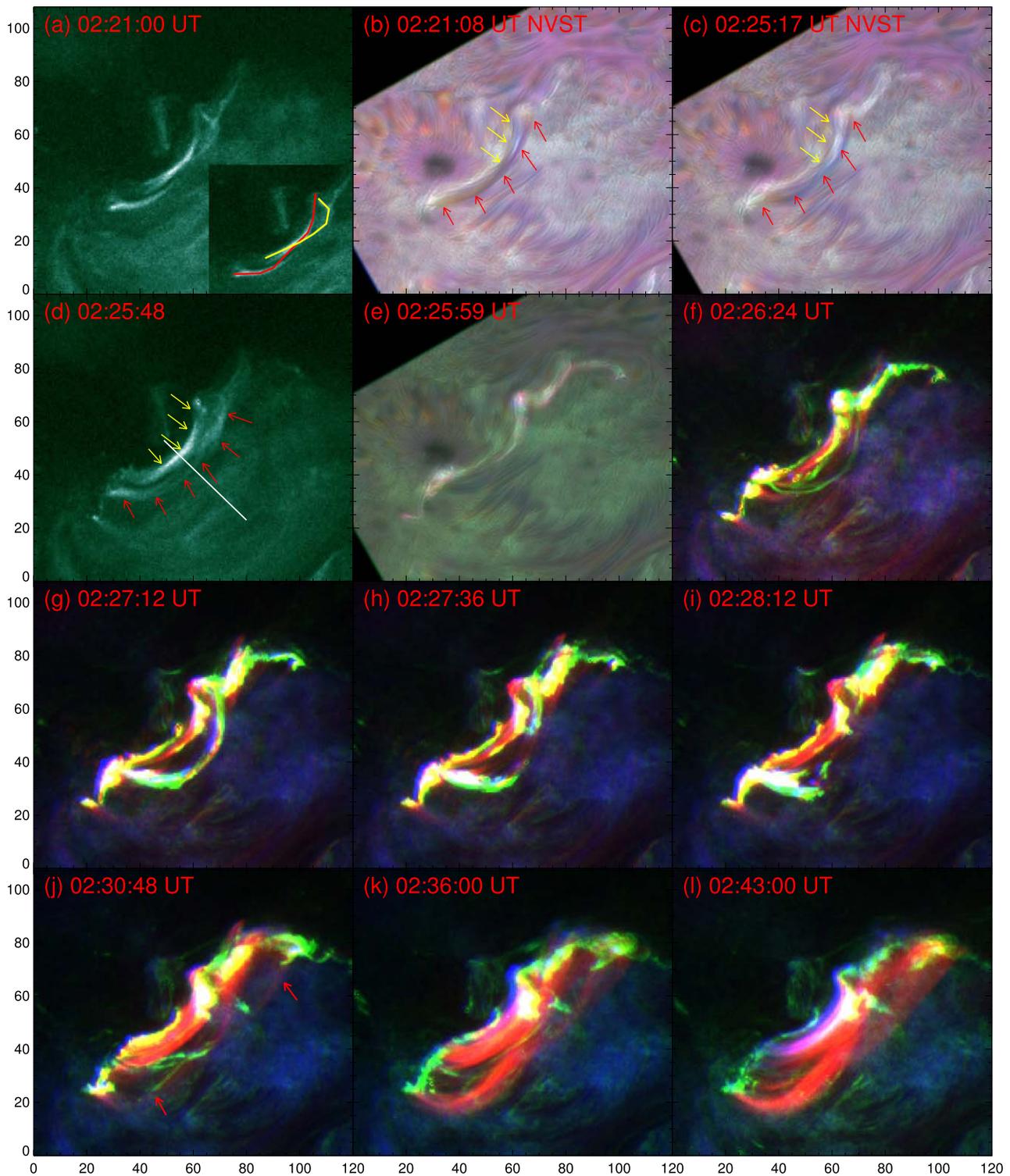

**Figure 4.** Dynamic evolution of the failed eruption of F1S. Panel (a) is an AIA 94 Å image. An enlarged image is shown on the bottom right, with red and yellow curves outlining the profile of the brightening threads. Panels (b), (c), and (e) are NVST images, with yellow arrows indicating the F3, and red arrows indicating the F1S. They show the slow rising of F1S. Panel (d) is an AIA 094 Å image, with yellow and red arrows outlining the profile of two parallel brightening threads. Panels (f)–(l) are composite images, with red for 94 Å, blue for 304 Å, and green for 211 Å.

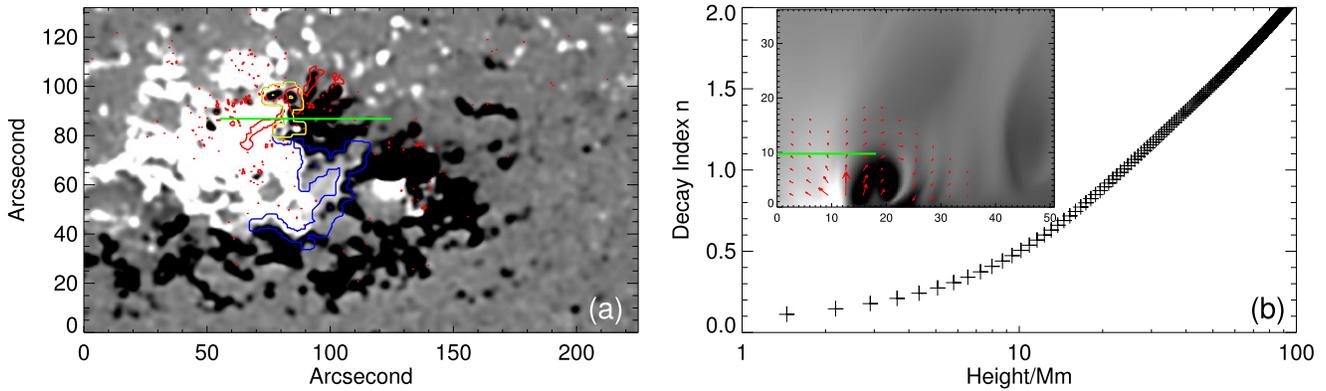

**Figure 5.** (a) Line-of-sight magnetic flied of AR 13102, with red contours outlining the flare region around the peak time, a blue contour for the PIL, and a yellow contour for the PIL within the core region of AR 13102. (b) The height profile of the decay index above the PIL region as outlined by the yellow contour in panel (a). The subimage is a snapshot of the NLFFF extrapolated magnetic field in the X–Z plane along the green line in Figure 5(a). The unit along the X- and Y-direction is megameters.

### 3.3. Plasma Flow

We show an overview of the plasma flow in Figure 6(a) with the AIA 171 Å image. The plasma flow appears as a bright structure in all cool AIA channels, most clearly at 171 Å with the temperature of 0.7 MK. In order to show the plasma flow track in one image, the maximum value image is used. The observational time periods for four subregions are labeled. The dynamic evolution of the plasma flow can be seen clearly by the animation attached to Figure 6 with running different images. As shown in Figure 6(a), the plasma flow originates from the recurrent reconnection region and moves toward a far remote area intermittently. The green curve indicates the main track of the plasma flow. Figure 6(b) shows the spatiotemporal evolution of the plasma flow seen through the green curve. The positions of three green stars in Figure 6(a) are corresponding to the thick short lines in Figure 6(b), which give referee positions for reading Figure 6(b). It is worth to note that Figure 6(a) is formed by four subregions, which are recorded by AIA at four special time periods. It is natural that the track of the plasma flow observed at other time period appears different from the one as we show in Figure 6(a). Meanwhile, the plasma flow tracked by the slit is its main part. As shown in Figure 6(b), before the filament eruption, the plasma flow originates from the recurrent reconnection region (around D1) and moves toward D2, and then becomes invisible in EUV observations. Associated with the filament eruption, first, the track from D2 to D3 starts brightening simultaneously. It means that the plasma continuously flows from D2 to D3, even though it is too weak to be identified in EUV observations. Second, after the failed filament eruption, the slanted brightening features, which represent the track of the plasma flow, appear more frequently and intensively in the area from D2 to D3 than those in the area from the original reconnection region to D2. Third, the track of most plasma flow is ended around D3. Meanwhile, the one that starts from about 02:36 UT moves toward a remote area and curves around a loop-like structure as shown in Figure 6(a). As shown in Figure 6(b), the sloping brightening features, which represent the features of the plasma flow, are dispersive, weak, and intermittent. We use a simple image segmentation method based on an edge enhancement algorithm (Sobotka et al. 1997) to bring out the skeleton of the plasma flow. Even though the plasma flow is still intermittent, some parts of them become easy to identify. An example is shown in the top right corner in Figure 6(b). This area is outlined by a green box. Here, the plasma flow is isolated as a brightening structure with an obvious edge feature. This edge is labeled by green dots by hand. In order to do the quantitative estimation of the velocity, a linear fitting is used. The red line shows the fitting result. The standard deviation of the position is about 28 km. The velocity deduced from the position is $94 \pm 2$ km s$^{-1}$. We do the same calculation for the other two features as indicated by blue arrows. The standard deviation of the position is about 120 and 230 km, with the velocity of $96 \pm 10$ and $119 \pm 19$ km s$^{-1}$, respectively.

We show two examples in Figure 7 for the plasma flows before the filament eruption. The first one (Figures 7(a)–(c)) is well recorded by NVST, and the second one (Figures 7(d)–(i)) shows difference evolution features in NVST and AIA. In Figures 7(a)–(c), the track of the plasma flow is indicated by blue arrows. As shown in these images, the plasma flow can be identified as an intermittent dark feature in a Hα image, and intermittent dark and brightening feature in EUV images. In Figure 7(a), an EUV brightening, which appears as the indirect observational evidence of magnetic reconnection, is outlined by a red contour. It is located around the intersection of the east foot-point of F2, the north foot-point of F1S, and the south foot-point of F1N. The east leg of F2 is circled around above the brightening. The image also shows that F2 is located above F1 and crosses F1 as an arch-shaped structure.

The images in Figure 7(d) show the early step of the second example with 304 Å image. The EUV brightening is also overlaid on the Hα image as a red contour in Figure 7(e). The blue arrows at the same place, as shown in Figures 7(a)–(c), are also shown in Figures 7(d) and (e). It means that the plasma flows at the difference time share the same channel. As shown in a later image of 94 Å (Figure 7(f)), F2 appears brightening. This brightening first appears around 02:04 UT. Figure 7(g) is a composite image, with red for Hα image and green for 94 Å image. It shows the consistency of the brightening of F2 and the plasma flow. Figures 7(h) and (i) are maximum value images during the time period from 02:00 UT to 02:10 UT, and from 02:20 UT to 02:35 UT, respectively. In order to show the fine structure of the flare core region, the area on the left of the first blue line in Figures 7(h) and (i) is formed by the maximum

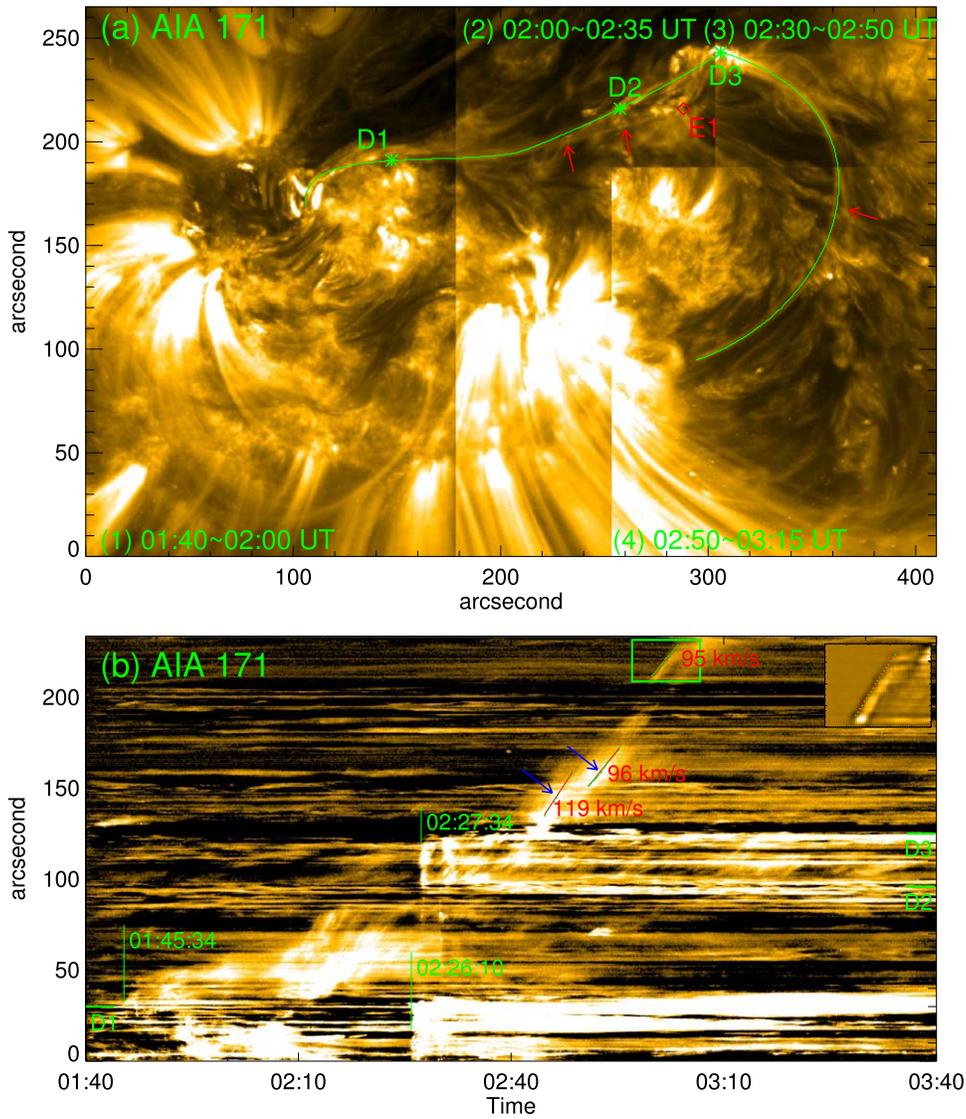

**Figure 6.** The AIA 171 Å observations to show the plasma flow clearly. (a) The maximum value image, composed by four subimages, with the time period is labeled. The green curve outlines the skeleton of the plasma flow. (b) Time-slice plot acquired along the green curve as shown in panel (a). The positions of three green stars in panel (a) are corresponding to the thick short lines in panel (b). The green box outlines the area we show in an enlarged image in the top right corner. The enlarged image is processed by a simple image segmentation method to show the skeleton of the plasma flow with obvious edge feature. The green dots outline the edge feature. The red line shows the linear fitting of this position. The blue arrows indicate two plasma flow features, which have been processed by the same method. An animation is available online to show the temporal evolution of the plasma flow. The animation begins at approximately September 19 01:40 UT and ends at around 03:40. The time cadence of the animation is about 24 s. The real-time duration of the animation is 26 s.

(An animation of this figure is available in the online article.)

value during the time period from 02:00 to 02:04 UT. In Figure 7(i), the area between two blue lines is formed by the maximum value during the time period from 02:04 UT to 02:10 UT. Red arrows in Figure 6(h) indicate the track of the plasma flow during time period from 02:00 to 02:10 UT. Its destination is labeled as a diamond. The diamond is also plotted at the same position in Figure 7(i). In the later phase as shown in Figure 7(i), the plasma track is different as the previous one.

Figure 8 shows the plasma flow associated with the failed filament eruption three times for examples. The subregion images in Figures 8(a) and (b) show the reappearance of the plasma flow simultaneously around the flaring time. In Figures 8(a) and (b), blue and red pluses mark two points. Around 02:27:34 UT, a brightening feature appears in between these two points, which is corresponding to the sudden appearance of the vertical brightening features in Figure 3(b).

Figure 8(c) shows the temporal evolution of the intensity of both points. It shows that both points become a brightening at the same time. The distance between them is about $2.0 \times 10^4$ km. It means that the plasma is already there and then is heated simultaneously.

Figure 8(d) shows the track of the plasma flow during the time period from 02:35 to 02:45 UT. The subimage at the bottom of Figure 8(d) shows two images, recorded around 02:37:46 and 02:43:46 UT. It shows that, during the time period from 02:27 to 02:39 UT, the destination of the plasma is D3. After that, the plasma flow makes a slow arc before reaching D3 to move toward the south. As shown in Figure 8(e), the plasma flow is composed of several threads. As shown in Figure 8(f), the plasma flow draws two trajectories. The one is a loop-like structure, and the other is too weak to be identified in its later phase.

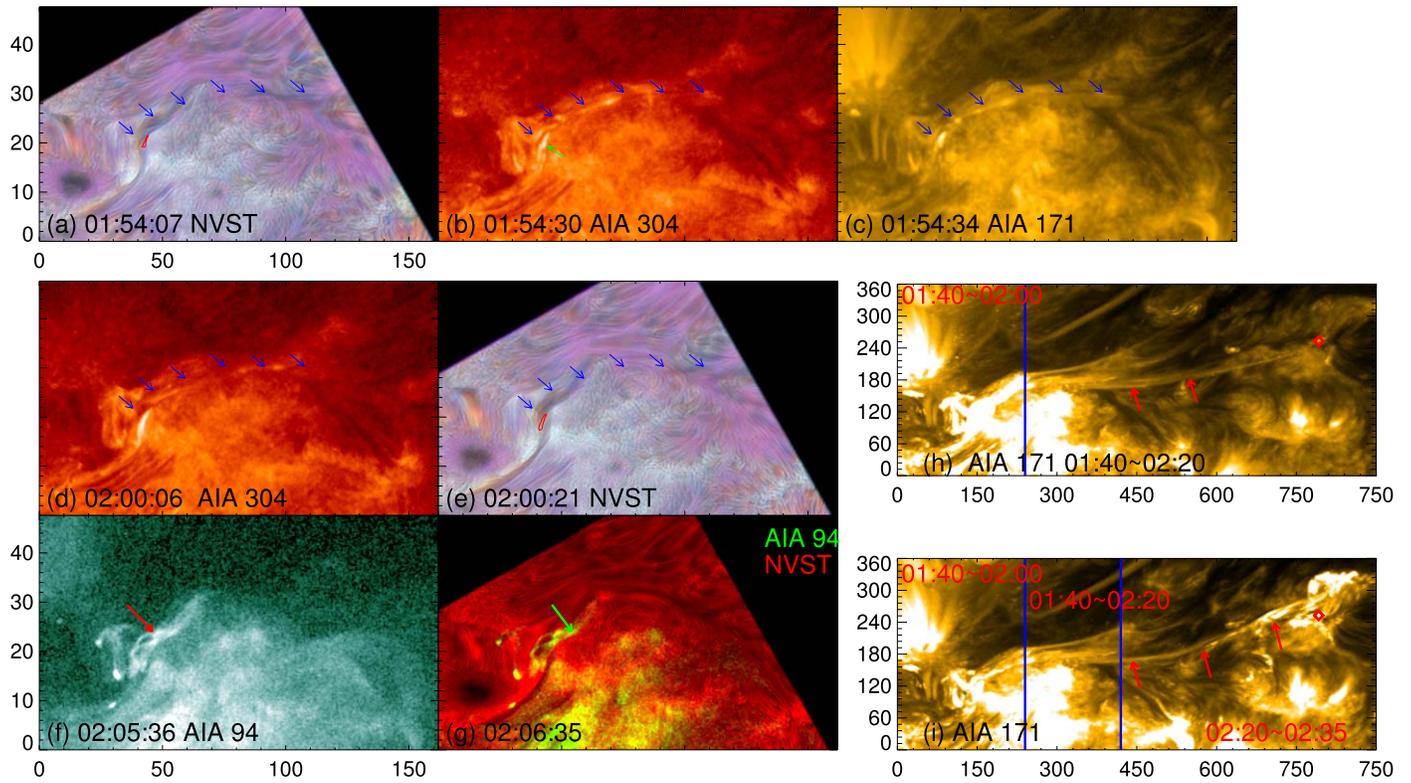

**Figure 7.** Dynamic temporal evolution of the plasma flow. Panels (a) and (e) are the NVST image. Panels (b) and (d) is the AIA 304 Å image. Panel (c) is the AIA 171 Å image, and panel (f) is the AIA 094 Å image. Panels (h) and (i) are a maximum value image of AIA 171 Å. Blue and red arrows outline the track of the plasma flow. The red arrow in panel (f) and green arrow in panel (g) indicate the brightening of F2. The red contour in panel (a) outlines the brightening, recorded in the AIA 304 Å image.

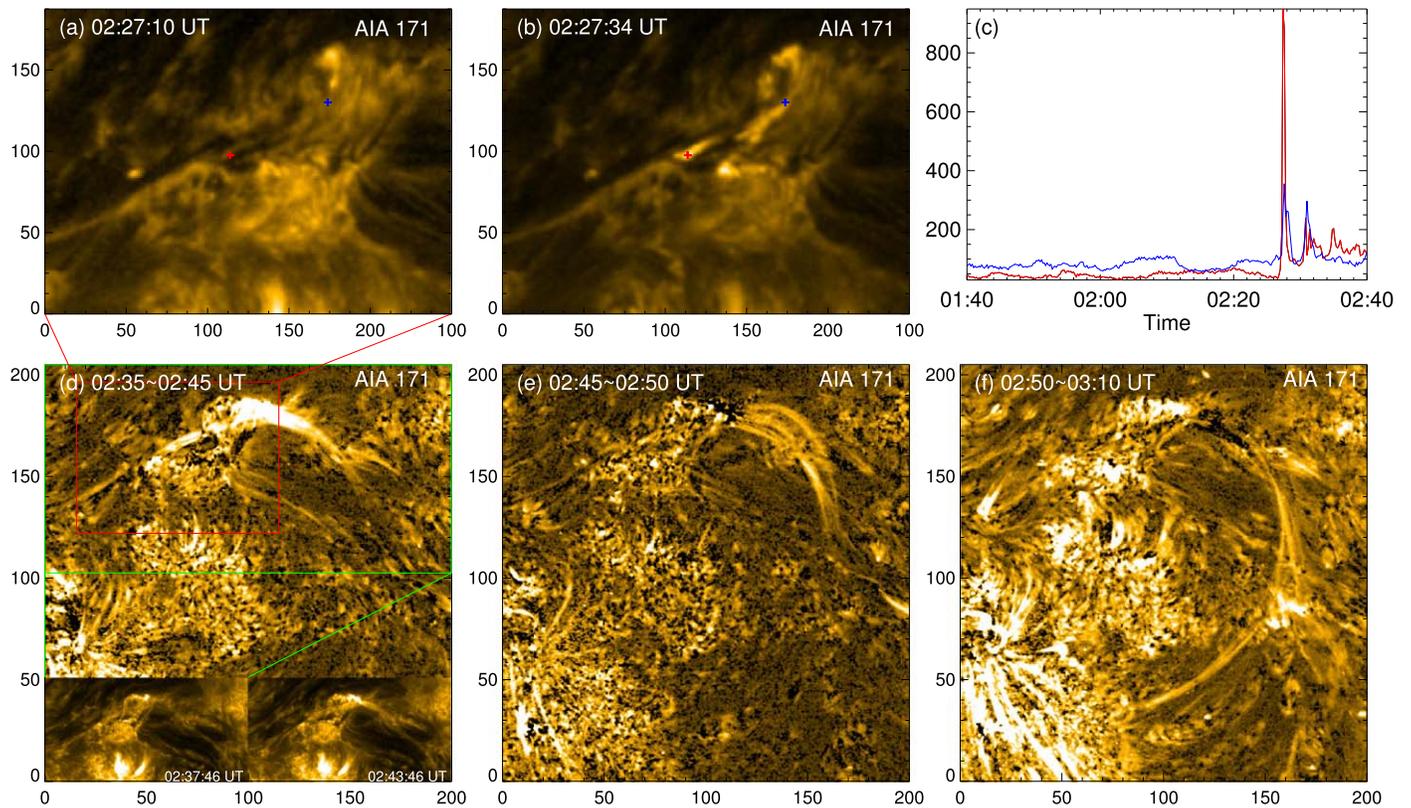

**Figure 8.** Dynamic evolution of the plasma flow associated with the failed filament eruption. Panels (a) and (b) show the sudden brightening of the plasma flow in a remote area associated with the failed filament eruption. Panel (c) shows the intensity evolution of two places, outlined by a red plus in panels (a) and (b). Panels (d)–(f) are the maximum value image, which shows the track of the plasma flow at different time period. More detail can be seen in the same animation with Figure 6.

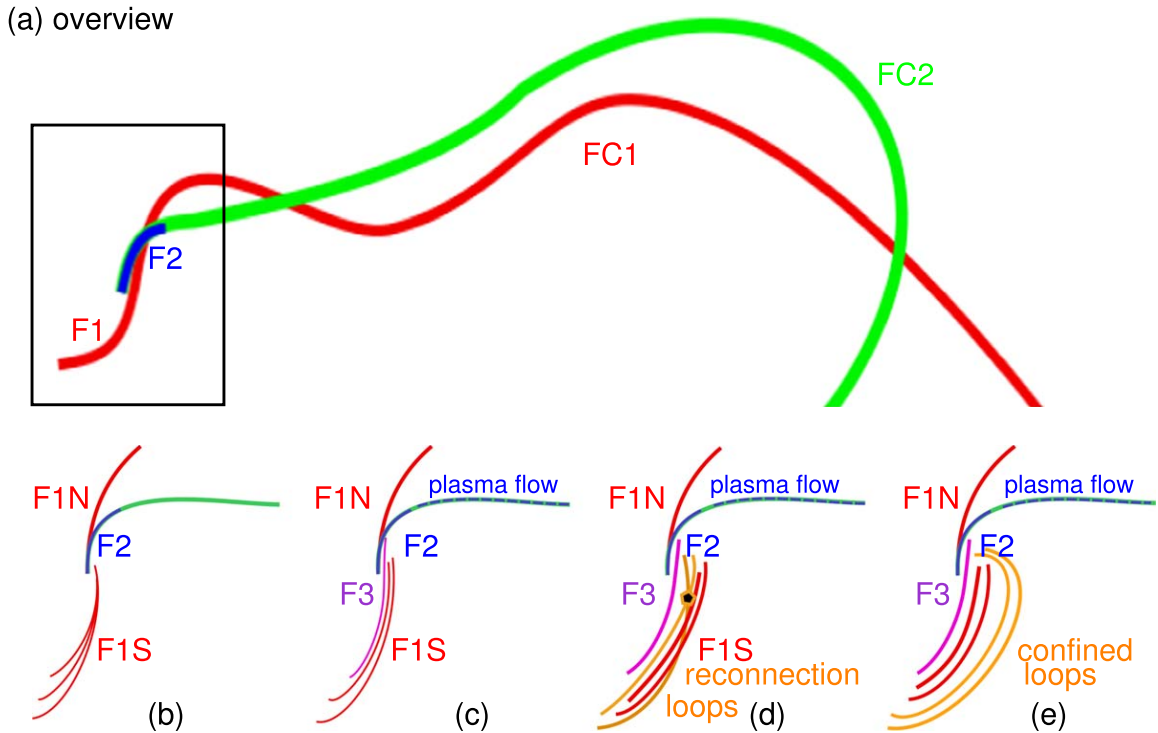

**Figure 9.** Schematic depictions of the proposed physical scenario. (a) Overview of the morphology of filaments and FCs. The black box outlines the east part, located within the core region of AR 13102. (b) Within the core region, the S-shaped filament is formed by F1N and F1S. The invisible FC2 is presented by a small filament of F2. (c) The new filament F3 is formed by the interaction between F1N and F1S. Associated with the recurrent reconnection, hot plasma is heated and flows along the FC2. The FC2 appears gradually associated with the plasma flow. (d) Reconnection occurs between the brightening coronal loops and pushes the remained F1S outward. (e) The outward-moving F1S is confined by the up coronal loops.

## 4. Conclusions

Based on the UV/EUV observations from AIA and SUTRI, and Hα images recorded by HSOS and NVST, we report complex activities of a double-decker filament and its host FCs on 2022 September 19. The activities involve small-scale reconnection, associated FC appearance, a C6.5 flare, and associated failed filament eruption. We summarize the observational results and proposed a physical scenario (Figure 9) as follows.

A long and intermittent filament, which winds through the core region of AR 13102 and its nearby quiet region along the PIL, is hosted by FC1 (Figure 9(a)). As seen from the high-spatial-resolution NVST image, within the AR, the filament section, which labeled as F1, is composed of two J-shaped structures (labeled as F1S/F1N for the south/north one) and appears as an S-shaped filament (Figure 9(b)). Recurrent reconnections are identified around the junction of F1S and F1N and result in the formation of a new long filament (F3; Figure 9(c)). Meanwhile, associated with the recurrent small-scale reconnection, plasma flow is identified. Its track is above the F1 and draws up a total different profile. Taking the plasma flow as reference object, another filament structure (F2) is identified, which is overlapped with F1N and only can be identified during the plasma flow process (Figure 9(a)). The intensity of the plasma flow declines with the distance between the reconnection site and the end of the plasma flow, and then becomes invisible around a remote area labeled as D2. The failed eruption of F1S is triggered by the reconnection, which appears as the brightening threads change their shape from crossed to quasiparallel in between the F3 and F1S (Figures 9(d) and (e)). Then, the F1S rises suddenly, and disappears in Hα. In EUV wavelengths, the F1S is always visible with obvious shrinkage. The decay index at the possible apex of the filament is 0.47. It is much smaller than the critical value of about 1.5. Associated with the F1S eruption, the invisible plasma flow suddenly becomes visible as a brightening feature from D2 to D3. And then continuous flows are from D2 to the remote foot-point of FC2. The plasma flow makes the empty FC2 appear gradually. The flowed plasma does not reach its remote foot-point, and the FC2 disappears associated with the dissipation of the hot plasma from D3 to the remote end.

## Acknowledgments

We thank the anonymous referee for the helpful comments. The authors thank the staff in Huairou observational station of National Astronomical Observatories in China for helpful discussion. Data were made available courtesy of NASA/SDO and the AIA science teams. The EUV images are also supplied by the SUTRI team. The ground-based Hα data are supplied by HSOS and NVST team. This research is supported by National Key R&D Program of China 2022YFF0503800 and 2021YFA1600500; Yunnan Key Laboratory of Solar Physics and Space Science (202205AG070009); NSFC grants 12373058, 12173050, 12003048, 12373115, and 12203097; Beijing Natural Science Foundation (grant No. 1222029); the Strategic Priority Research Program of the Chinese Academy of Sciences XDB0560302; and the International Partnership Program of Chinese Academy of Sciences (183311KYSB20200003).

## ORCID iDs

Yin Zhang https://orcid.org/0000-0002-0093-0350
Baolin Tan https://orcid.org/0000-0003-2047-9664
Quan Wang https://orcid.org/0000-0003-3142-217X
Jing Huang https://orcid.org/0000-0001-8250-1535
Zhe Xu https://orcid.org/0000-0002-9121-9686
Kanfan Ji https://orcid.org/0000-0001-8950-3875
Xiao Yang https://orcid.org/0000-0003-1675-1995
Jie Chen https://orcid.org/0000-0001-7472-5539
Xianyong Bai https://orcid.org/0000-0003-2686-9153
Zhenyong Hou https://orcid.org/0000-0003-4804-5673
Yuanyong Deng https://orcid.org/0000-0003-1988-4574